\documentclass{PoS}

\usepackage{subfigure}
\usepackage{url}
\def\lsim{\mathrel{\rlap{\lower4pt\hbox{\hskip1pt$\sim$}}
    \raise1pt\hbox{$<$}}}         
\def\gsim{\mathrel{\rlap{\lower4pt\hbox{\hskip1pt$\sim$}}
    \raise1pt\hbox{$>$}}}         

\title{Observation of Coherent Elastic Neutrino-Nucleus Scattering by COHERENT}

\ShortTitle{COHERENT}

\author{\speaker{Kate Scholberg, for the COHERENT collaboration}\\
        Department of Physics, Duke University, Durham, NC 27708, USA\\
        E-mail: \email{schol@phy.duke.edu}}

\abstract{The COHERENT collaboration measured coherent elastic neutrino-nucleus scattering (CEvNS) for the first time at the Spallation Neutron Source at Oak Ridge National Laboratory, using a CsI[Na] detector~\cite{Akimov:2017ade}.  Here we discuss the nature of the CEvNS process, physics motivations, and experimental considerations for measuring CEvNS.  We describe the CsI[Na] measurement, along with status and future of COHERENT.}

\FullConference{The 19th International Workshop on Neutrinos from Accelerators-NUFACT2017\\
		25-30 September, 2017\\
		Uppsala University, Uppsala, Sweden}

\begin{document}

\section{Introduction: CEvNS}

The neutral-current (NC) coherent elastic neutrino-nucleus scattering (CEvNS) process results when a neutrino interacts with a nucleus via exchange of a Z boson, and the nucleus recoils as a whole.  The process dominates for medium-$A$ nuclei for neutrino energies less than about 50~MeV.  The coherence condition corresponds to target nucleons in phase with each other at low momentum transfer, $Q$, for $QR<<1$, where $R$ is the size of the nucleus.  The hallmark relation for a coherent interaction is that the total cross section scales as the square of the number of constituents times the single constituent cross section.\footnote{Note that a neutrino-interaction process may be coherent but not elastic; an example of an \textit{inelastic} neutrino-nucleus process is coherent pion production, recently observed~\cite{Higuera:2014kq}; however, this $\pi$-production process produces new particles and is distinct from CEvNS, which just involves the recoil of a nucleus.}.

The existence of CEvNS was first proposed 43 years ago~\cite{PhysRevD.9.1389}, and its detection was called an ``act of hubris'' due to ``grave experimental difficulties.''   
The cross section has a straightforward Standard-Model (SM) prediction~\cite{Freedman:1977xn}:

  \begin{eqnarray}\label{eq:sevens}
\frac{d\sigma}{dT}_{coh} &=& \frac{G_F^2 M}{2\pi}\left[(G_V + G_A)^2 + (G_V - G_A)^2\left(1-\frac{T}{E_{\nu}}\right)^2 - (G_V^2 - G_A^2)\frac{MT}{E_{\nu}^2}\right] \\
G_V &=& (g_V^p Z + g_V^n N)F_{\rm nucl}^V(Q^2)\\
G_A &=& (g_A^p(Z_+ - Z_-) + g_A^n(N_+ - N_-))F_{\rm nucl}^A(Q^2),
\end{eqnarray}
where $G_F$ is the Fermi constant, $M$ is the nuclear mass, $T$ is the recoil energy, $E_{\nu}$  neutrino energy, 
$g_V^{n,p}$ and $g_A^{n,p}$ are the SM vector and axial-vector coupling factors, respectively, for protons and neutrons, 
$Z$ and $N$ are the proton and neutron numbers, $Z_{\pm}$ and $N_{\pm}$ refer to the number of up or down nucleons, and $Q$ is the momentum transfer~\cite{Barranco:2005yy}. 

The vector contribution dominates strongly for most nuclei; because the axial contributions depend on unpaired nucleons, for which numbers are typically much smaller than total numbers of nucleons, axial contributions are small, and they are zero for spin-zero nuclei.

The form factor $F(Q)$ encapsulates information about the nucleon distributions.  It suppresses the cross section at large $Q$.  Form factors are known to $\sim$5\% or better.

\begin{figure}[htb]
\begin{center}
\includegraphics[width=3.in]{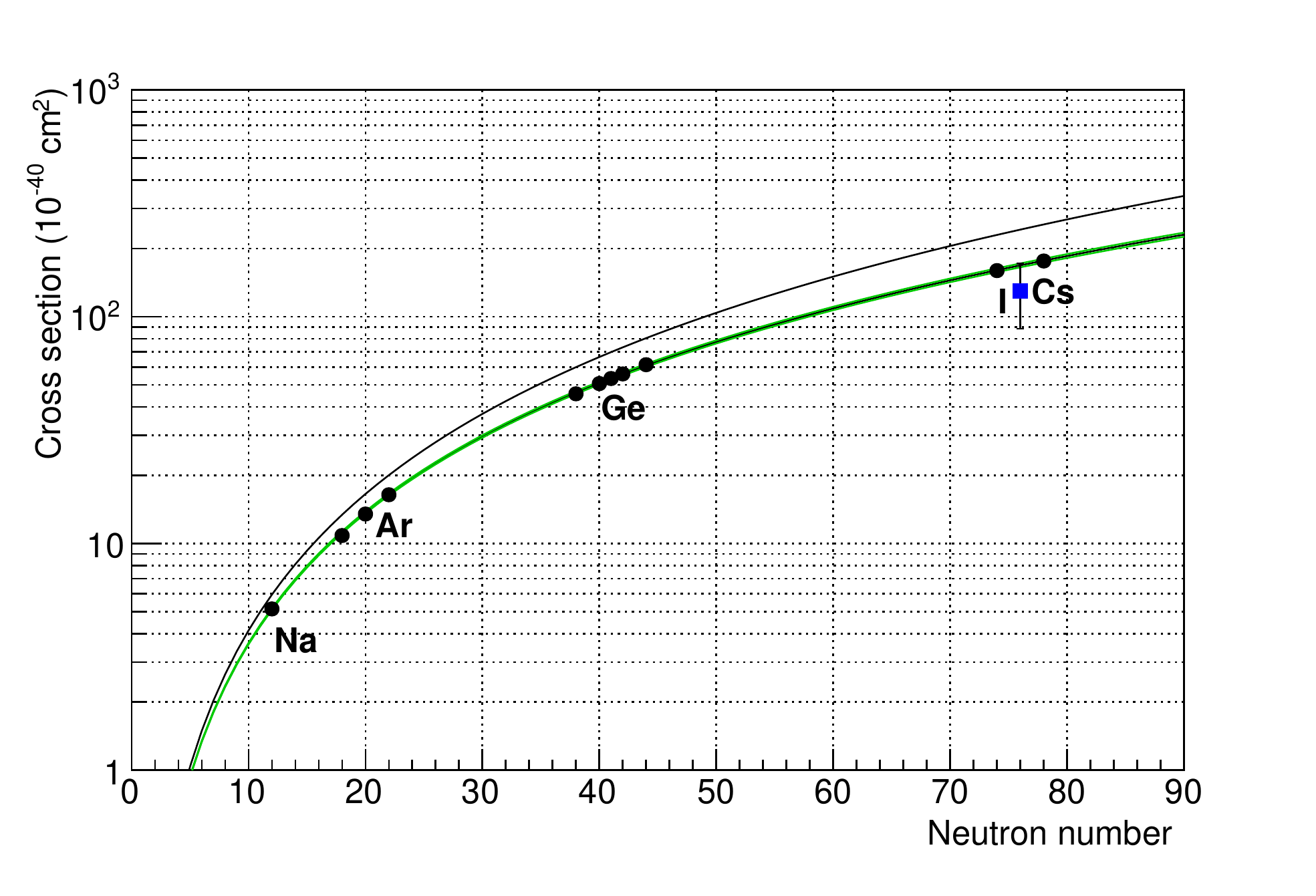}
\includegraphics[width=2.8in]{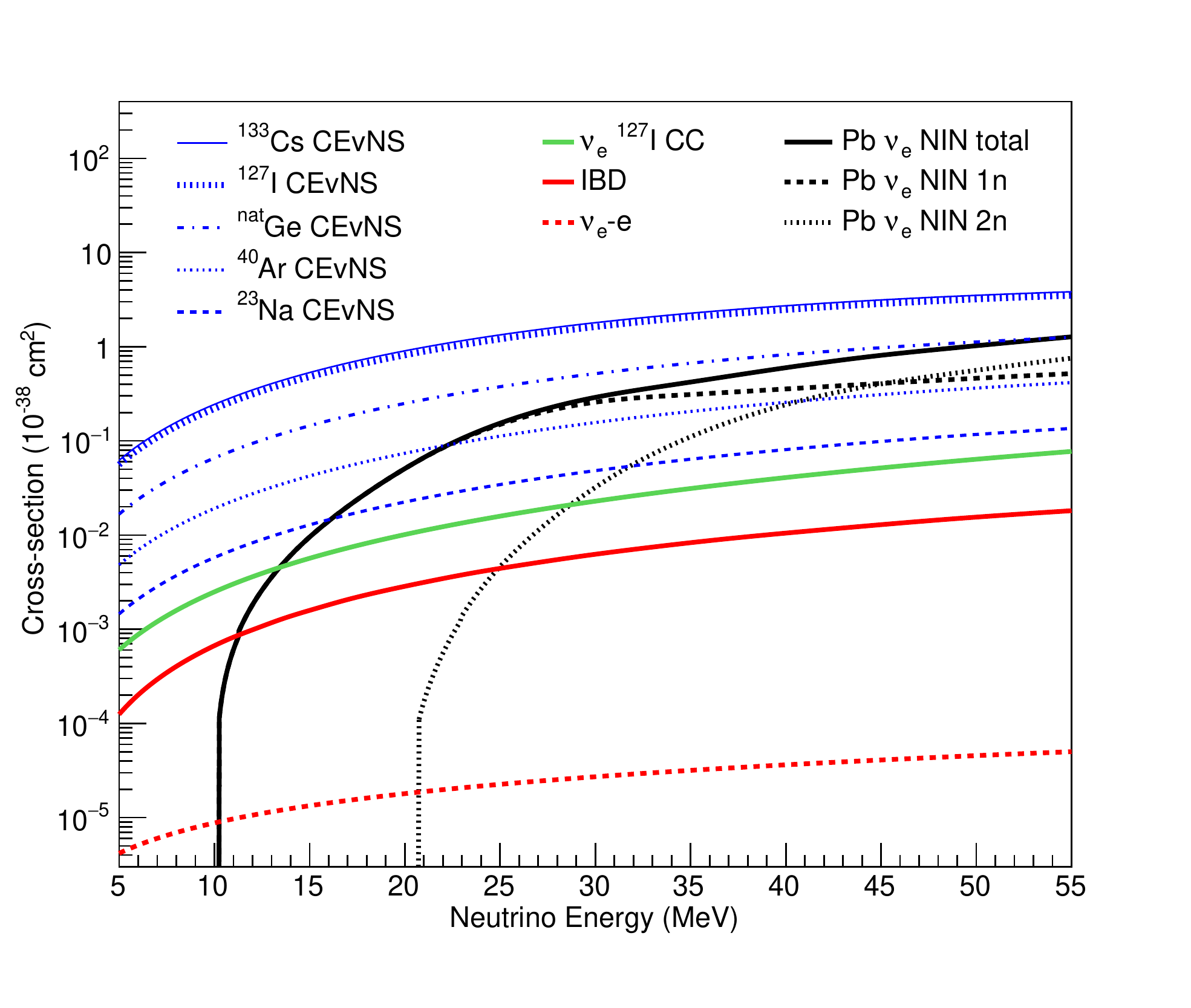}
\caption{\label{fig:xscns} 
Left: Illustration of the $\propto N^2$ proportionality of the stopped-pion-neutrino flux-averaged CEvNS cross section versus neutron number $N$. The black line assumes a unity form-factor. The green band shows the effect of an assumed Helm form factor~\cite{Helm:1956dv} with its width indicating the effect of a $\pm 3\%$ uncertainty on the assumed neutron rms radius.  The points show the the relevant isotopes of COHERENT target materials.  The blue square shows the flux-averaged cross section inferred from the measurement reported in Ref~\cite{Akimov:2017ade}.  Right: Neutrino interaction cross sections per target as a function of neutrino energy for COHERENT target materials, as well as NIN cross sections on lead.  Also shown
are inverse beta decay of $\bar{\nu}_e$ on free protons (IBD), and elastic scattering of $\nu_e$ on electrons (per electron).
}
\end{center}
\end{figure}

Neglecting axial terms and for $T<<E_\nu$, the differential recoil spectrum is approximately given by $\frac{d\sigma}{dT} \sim \frac{G_F^2 M}{2 \pi} \frac{Q_W^2}{4} F^2(Q)\left(2-\frac{M T}{E_{\nu}^2}\right)$.
The event rate is proportional to the weak nuclear charge, $Q_W=N-(1-4\sin^2\theta_W)Z$.  Since $\sin^2\theta_W=0.231$ is close to $1/4$, the contribution from the protons in the nucleus, proportional to $Z$, is unimportant compared to the $N$ contribution.  Therefore, $\frac{d\sigma}{dT} \propto N^2$.  Fig.~\ref{fig:xscns} illustrates this $N^2$ dependence, along with the form-factor suppression.  A measurement in multiple nuclear isotopes showing the expected $N^2$ dependence will confirm this expected property of the CEvNS interaction.

Thanks to the $N^2$ enhancement of the cross section, the cross section is relatively larger than other charged-current (CC) and NC interactions in this
energy range: see Fig.~\ref{fig:xscns}.  One might ask then why experimental difficulties are so grave~\cite{PhysRevD.9.1389}.   The reason is kinematics--- the nuclear recoil energies are tiny.  For example, for a 30-MeV neutrino, the maximum recoil energy for CEvNS with Ge is $\sim 2 E_\nu^2/M$, which is only about 25~keV, and well below the threshold of conventional neutrino detectors.   The only experimental signature that a CEvNS interaction has occurred is the tiny energy deposited by the nuclear recoil in the target material.  Fortunately, since the publication of \cite{PhysRevD.9.1389}, there has been much development of low-energy recoil detectors, motivated to large degree by the search for weakly interacting massive particles (WIMPs).

\section{Physics Motivation}

 Several motivations for a CEvNS measurement are listed below, with example references.

\begin{figure}[ht]
\centering
\includegraphics[width=2.6in]{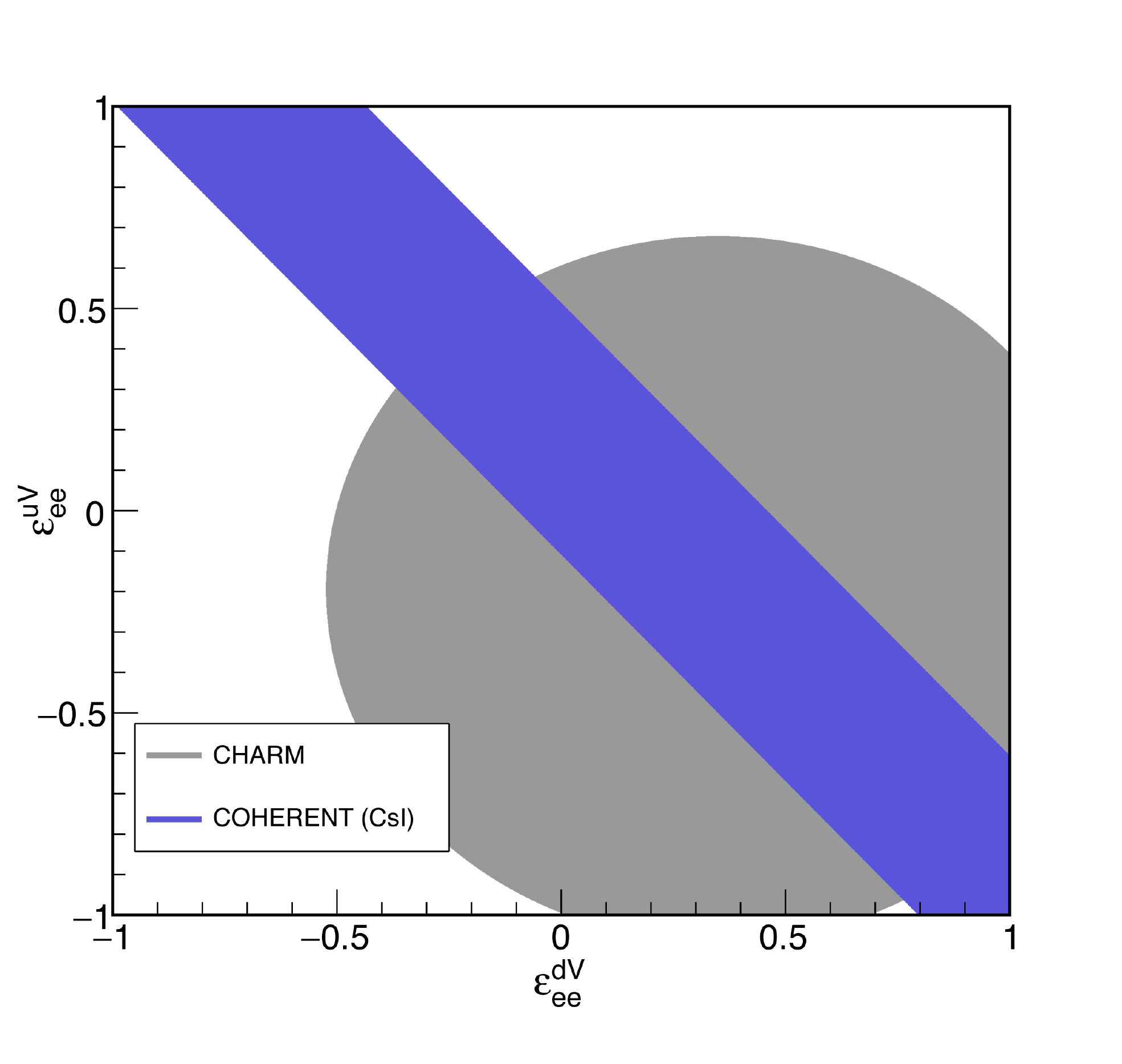}
\includegraphics[width=2.6in]{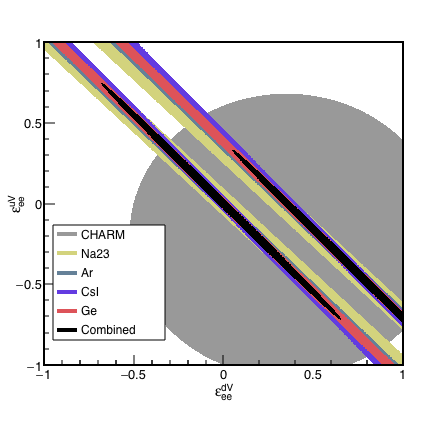}
\caption{\label{fig:NSI}Left: result from ~\cite{Akimov:2017ade} with initial constraints on two of the NSI $\epsilon$ parameters, showing also the constraint from the CHARM experiment {\protect\cite{Dorenbosch:1986tb}}.  
Right:
predicted sensitivity obtained with the COHERENT detector materials, with 5\% uncertainties on flux and event rates and assuming systematically-limited measurements. The black shows the result from a combined fit.   }
\end{figure}

\begin{itemize}

\item CEvNS represents a background for direct dark matter experiments; the so-called ``neutrino floor'' due to CEvNS interactions of solar, supernova and atmospheric neutrinos will eventually limit the sensitivity of next-generation WIMP detectors (e.g., ~\cite{billard:2014}).  This represents also a signal opportunity, since solar and supernova neutrinos are also accessible to large recoil-sensitive detectors (e.g.,~\cite{Billard:2014yka}).
\item Because the CEvNS cross section is a clean prediction of the standard model, a deviation would represent new physics.    Potential beyond-the-SM physics accessible to CEvNS experiments includes: non-standard interactions of neutrinos, new mediators, and large neutrino magnetic moment (e.g., ~\cite{Scholberg:2005qs}).  See Fig.~\ref{fig:NSI}.
In the context of the standard model, one can measure $\theta_W$ in a new way.
Several recent references explore these possibilities (e.g., ~\cite{Kosmas:2017tsq} and references therein).
\item Because the NC interaction is flavor blind, CEvNS is a new tool for sterile oscillation searches (e.g.,~\cite{Anderson:2012pn}).
\item CEvNS is important in supernova processes  (e.g.,~\cite{Sato:1975id}) and also allows detection of supernova burst neutrinos (e.g.,~\cite{Horowitz:2003cz})
\item Nuclear uncertainties in the CEvNS process prediction are less than 5\%, but a measurement with precision better than that can get at neutron form factors~\cite{Patton:2013nwa,Cadeddu:2017etk}, and be a probe of weak interactions in nuclei, with potential relevance to $g_A$ quenching.
\item The CEvNS cross section is large enough that it potentially has practical applications; reactor monitoring has been proposed (e.g.,~\cite{Hagmann:2004pb}). 

\end{itemize}

\section{CEvNS at a Stopped-Pion Source}

To measure CEvNS, one needs a neutrino source and a detector sensitive to low-energy recoils.  Desirables for the source are high flux, a well-understood spectrum, multiple flavors (good for BSM physics sensitivity, which may be flavor dependent), low background (for which a pulsed source is advantageous), as well as practical things such as ability to get close, good access, etc.

There have been many attempts over the years, and attempts are ongoing (e.g., \cite{Agnolet:2016zir, Aguilar-Arevalo:2016khx,conus,Belov:2015ufh}) to measure CEvNS at nuclear reactors, where neutrino energies go up  about 8~MeV.  The CEvNS cross section scales as the square of the neutrino energy; furthermore, the maximum recoil energy also scales as the 
square of the neutrino energy, $T_{\rm max} \sim 2 E_\nu^2/M$.  Therefore, for
best observability, one wants as large a neutrino energy as possible, while keeping
 violation of the coherence condition to a minimum, i.e., while maintaining $Q<1/R$.

\begin{figure}[ht]
\centering
\subfigure[SNS neutrino energy spectrum.]{%
\includegraphics[width=0.45\linewidth]{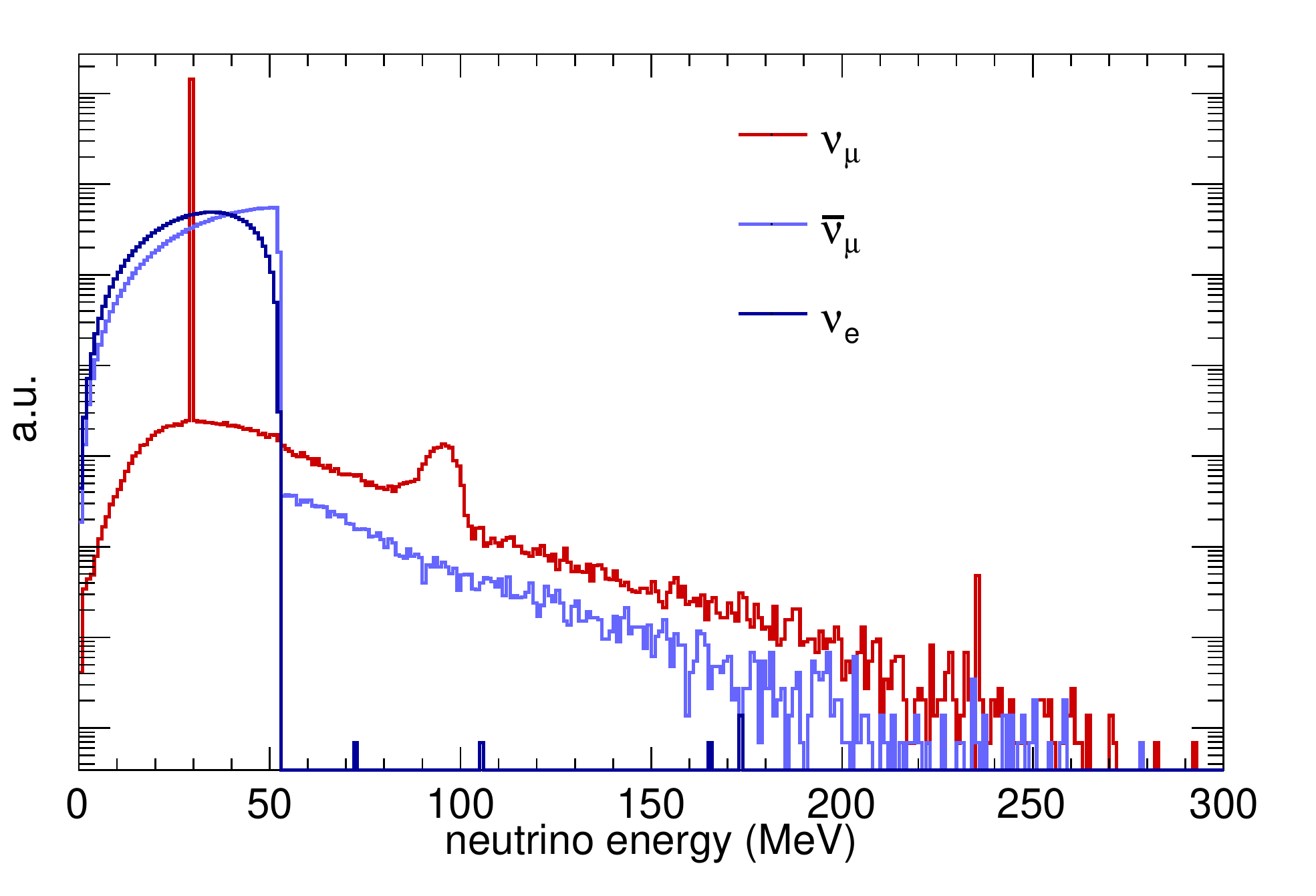}
\label{fig:snsisawesomea}}
\quad
\subfigure[SNS neutrino timing distribution.]{%
\includegraphics[width=0.45\linewidth]{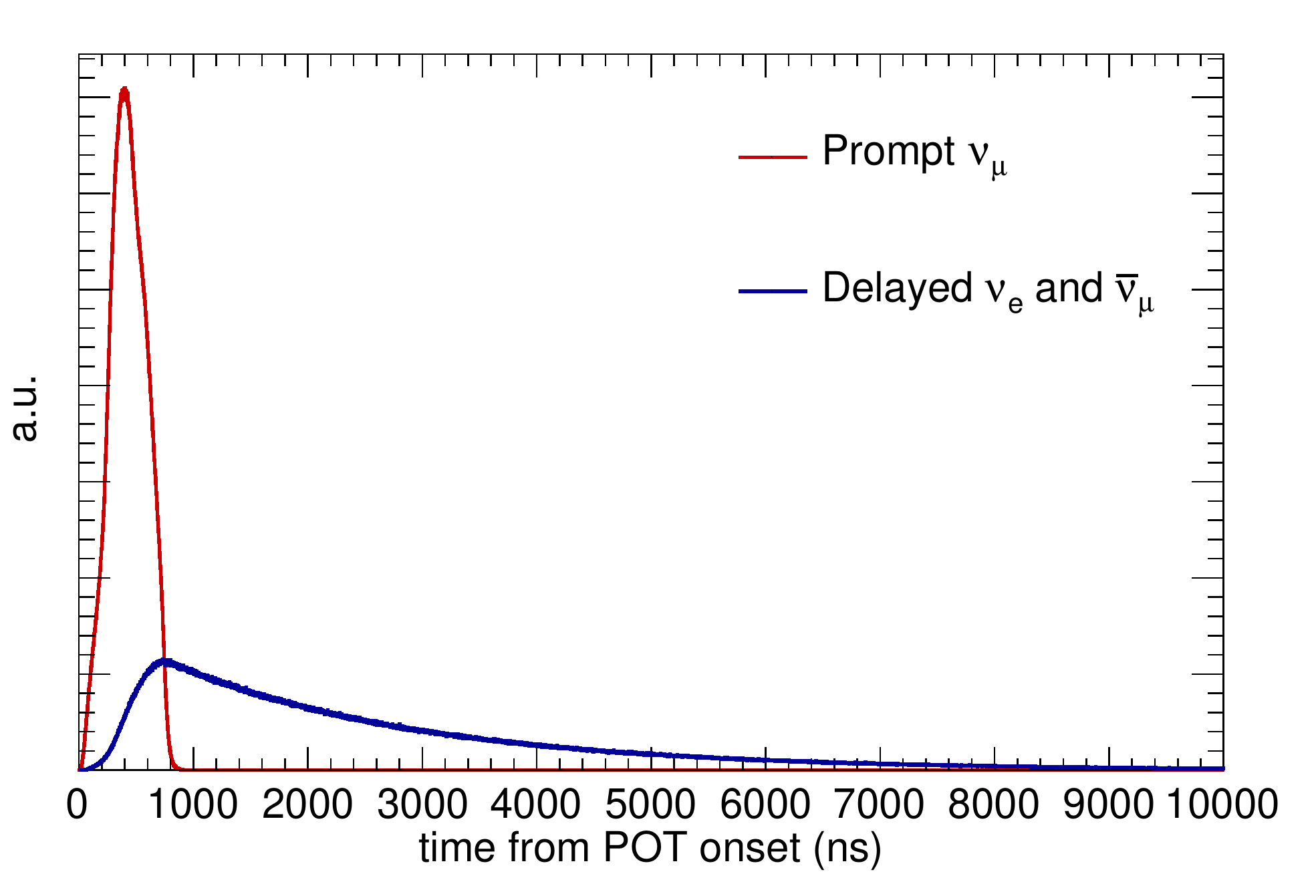}
\label{fig:snsisawesomeb}}
\caption{(a) Expected $\nu$ spectrum at the SNS, showing the very low level of decay-in-flight and other non-decay-at-rest flux, in arbitrary units; the integral is $4.3 \times 10^7$ neutrinos/cm$^2$/s at 20~m. (b) Time structure for prompt  and delayed neutrinos due to the 60-Hz pulses. ``Prompt'' refers to neutrinos from pion decay, and ``delayed'' refers to neutrinos from muon decay.}
\label{fig:snsisawesome2}
\end{figure}

Neutrinos from pion decay at rest, which have energies up to about 50 MeV, represent a sweet spot for CEvNS detection.  Pions are produced by $\sim$GeV protons incident on a target; one gets a prompt monochromatic 29.9-MeV neutrino from the pion decay, followed by $\bar{\nu}_\mu$ and $\nu_e$ on a 2.2-$\mu$s muon decay timescale with a well-understood 3-body-decay spectrum.  The Spallation Neutron Source (SNS) at Oak Ridge National Laboratory provides a very high-quality source of stopped-pion neutrinos, with favorable power ($\sim$MW) and beam time-profile properties (background rejection factor of $10^3-10^4$--- see Fig.~\ref{fig:snsisawesome2}).  Furthermore,  the $\sim$1-GeV proton energies are low enough, and the mercury target is large and dense enough, that nearly all pions decay at rest, resulting in a very clean, well-understood neutrino spectrum.

Steady-state backgrounds, such as intrinsic and ambient radioctivity and cosmogenics, can be suppressed strongly (and measured off beam pulse) by the pulsed nature of the source.  Neutron backgrounds in time with the beam cannot be suppressed in this way, and so must be carefully characterized and estimated. A ``friendly-fire'' in-time background results from neutrino-induced neutrons, or NINs, which are neutrons ejected from shielding materials such as lead, for which cross section is relatively high.  This background represents an interesting signal in itself, as it can be used for supernova detection~\cite{Duba:2008zz}.

\section{The COHERENT Experiment and First Light}

\begin{figure}[ht]
\centering
\includegraphics[height=2.7in]{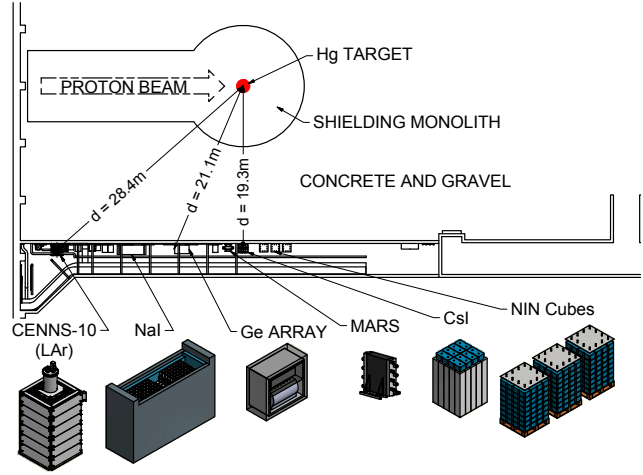}
\vspace{-0.2in}
\caption{Siting of existing and near-future planned detectors in Neutrino Alley. }\label{fig:neutrinoalley}

\end{figure}

\begin{figure}[ht]
\centering
\includegraphics[height=3.0in]{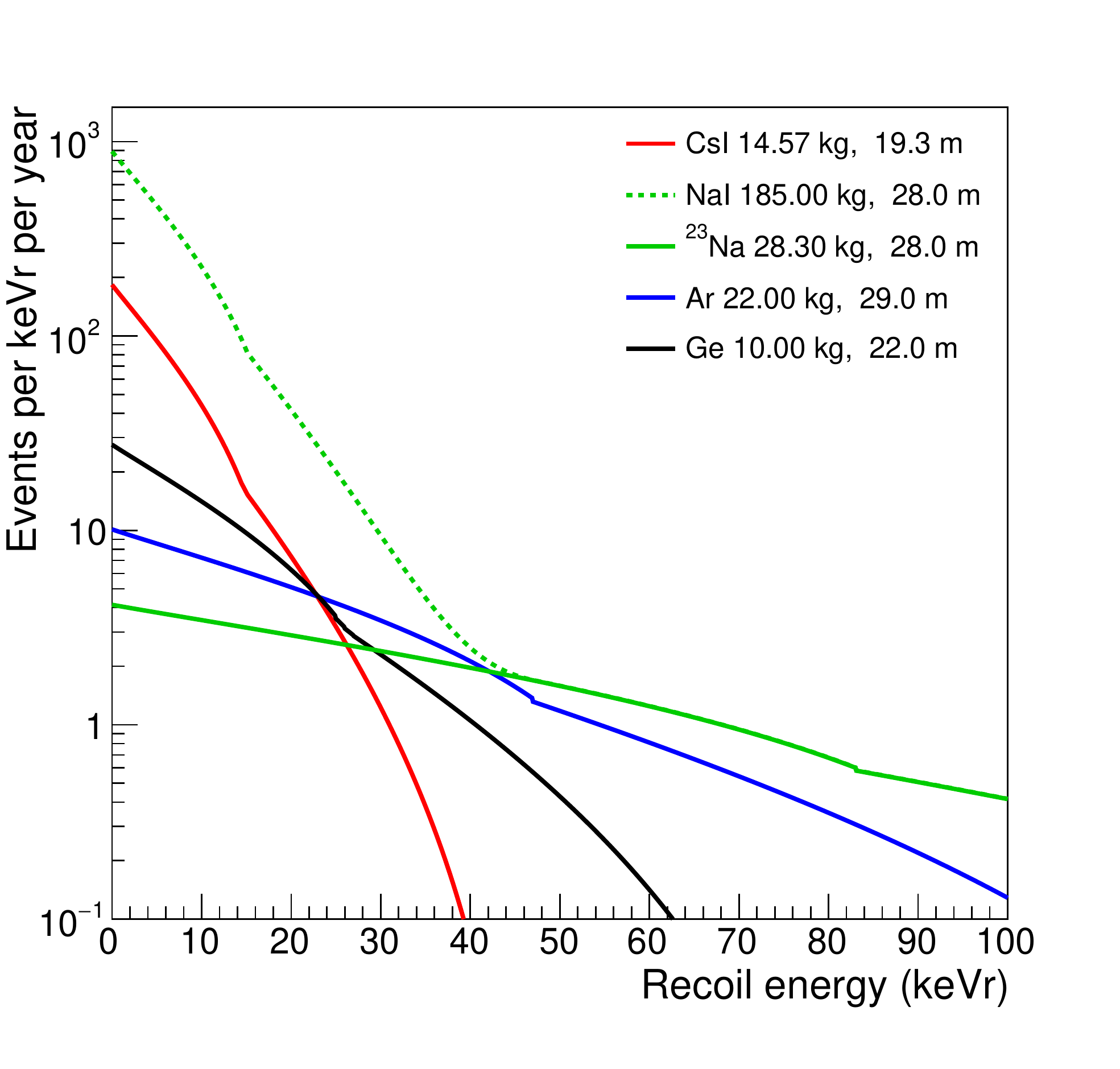}
\caption{Differential recoil rates for the COHERENT suite of detectors in Neutrino Alley.   This plot represents interaction rates for the first 6000 $\mu$s after each beam pulse, including all flavors and assuming 100\% detection efficiency.   Note this does not take into account quenching factors, which are different for different detectors, or other detector-specific efficiencies.  The kinks are due to different endpoints of prompt and delayed flavor components for different target isotope components.   The contribution from $^{23}$Na to NaI is shown separately, as the small quenching factor for $^{127}$I in NaI~\cite{Collar:2013gu}  is likely to strongly suppress the $^{127}$I contribution to the observed rate.  The materials are assumed to have natural abundances of isotopes.
}
\label{fig:spectrum}
\end{figure}

The COHERENT collaboration~\cite{Akimov:2015nza} aims to measure CEvNS in multiple target materials at the SNS.   The COHERENT detectors are sited in a basement hallway known as ``Neutrino Alley.''  The detectors are described in Tab.~\ref{tab:detectors}, and their siting is shown in Fig.~\ref{fig:neutrinoalley}.  The ``sanitized'' nuclear recoil spectra for these targets (i.e., no ``quenching'' or efficiencies applied) are in Fig.~\ref{fig:spectrum}.

\begin{table*}[htbp]
	\centering
 	\begin{tabular}{c|c|c|c|c}
		\hline
 		Nuclear& Technology & Mass & Distance from & Recoil \\

		target & & (kg) &  source (m) & threshold (keVnr)
		\\ \hline 

		CsI[Na] & Scintillating crystal & 14.6 & 19.3 & 6.5 \\ 
		Ge & HPGe PPC & 10 & 22 & 5 \\
		LAr & Single-phase & 22 & 29 & 20 \\
		NaI[Tl] & Scintillating crystal & 185$^*$/2000 & 28 & 13 \\

		\hline
	\end{tabular}
	\caption{\label{tab:detectors}Parameters for the four COHERENT
          detector subsystems. $^*$Deployed in high-threshold
          mode.}
 \end{table*}

\begin{figure}[ht]
\centering
\includegraphics[width=5.0in]{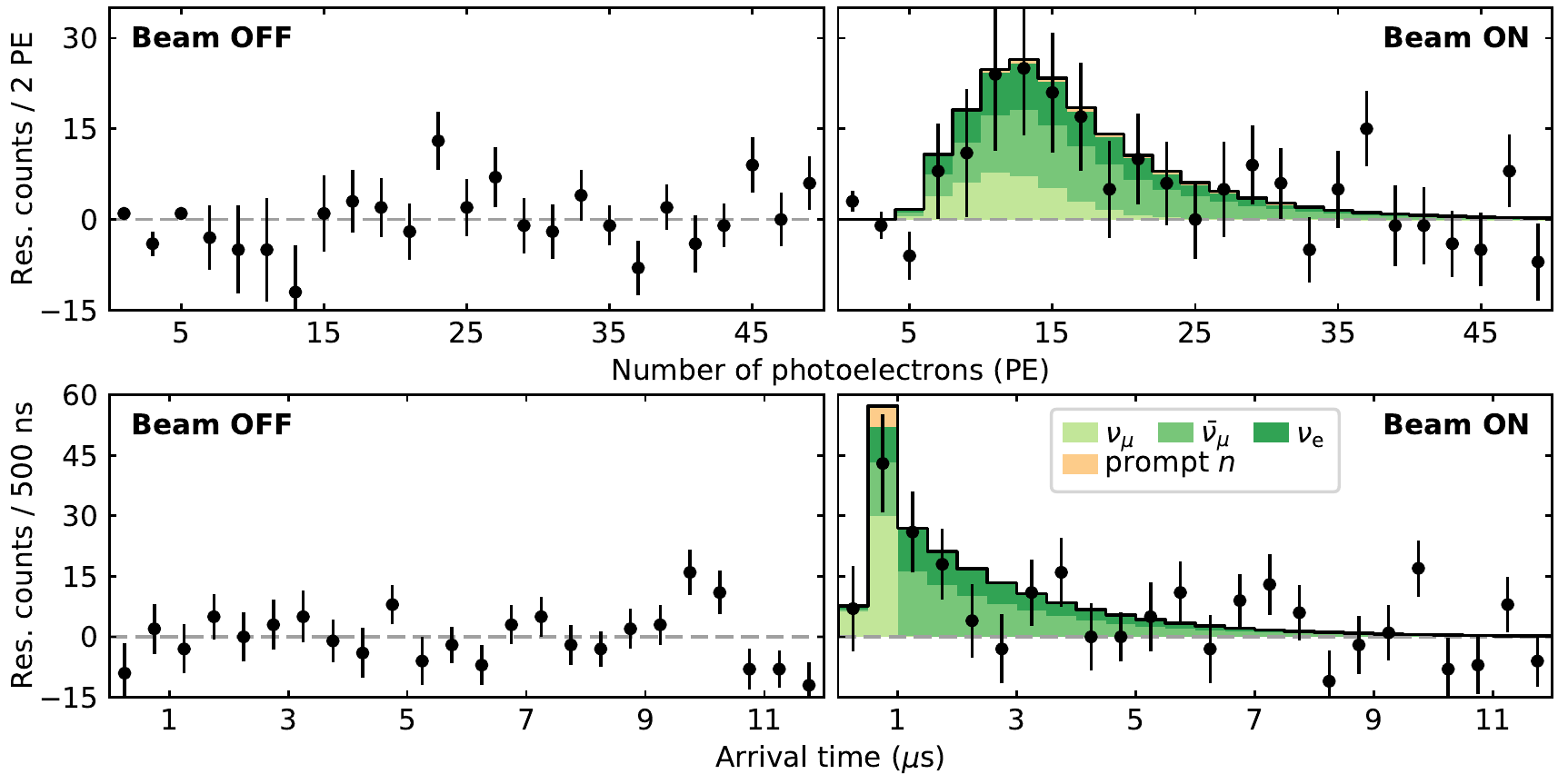}
\caption{Figure and caption from ~\cite{Akimov:2017ade}: residual differences (data points) between CsI[Na] signals in the 12 $\mu$s following POT triggers, and those in a 12-$\mu$s window before, as a function of their (top) energy (number of photoelectrons detected), and of (bottom) event arrival time (onset of scintillation). 
Error bars are statistical. These residuals are shown for 153.5 live-days of SNS inactivity (``Beam OFF'') and 308.1 live-days of neutrino production (``Beam ON''), over which 7.48~GWhr of energy ($\sim$1.76 $\times 10^{23}$ protons) was delivered to the mercury target.  
}
\label{fig:firstlight}
\end{figure}

The first measurement by the COHERENT collaboration was made using the 14.57-kg CsI[Na] crystal detector~\cite{Collar:2014lya}.    Data-taking started in summer 2015, and the first result was based on $1.76 \times 10^{23}$ protons on target (7.48~GWhr of exposure).  The charge and time distributions (residual counts subtracting anticoincidence data) are shown in Fig.~\ref{fig:firstlight}, and compared to beam-off samples.  The best fit to the data is $134\pm 22$ observed events, rejecting no signal at 6.7$\sigma$.   This result is consistent with the SM prediction of 173 events, with a $\sim$28\% uncertainty.  The uncertainty on the prediction is dominated by the uncertainty on the quenching factor (detector response) of CsI[Na].   Flux uncertainty is estimated at 10\%, and event-selection and form-factor uncertainties at 5\% each.

The interpretation of this result as a limit on NSI interactions is shown in Fig.~\ref{fig:NSI}.  Several authors have made further interpretation constraining BSM physics, e.g., ~\cite{Coloma:2017ncl}.

\section{Status and Future}

The immediate future plans for the existing COHERENT detector subsystems are:

\begin{itemize}

\item CsI will continue running.
\item An NaI[Tl] detector of 185 kg mass was installed in 2016, and is currently taking data in high-threshold mode, sensitive to CC interactions on $^{127}$I.  It will be refurbished for lower threshold to enable CEvNS sensitivity.
\item A LAr single-phase detector was installed in 2016 and upgraded in May 2017.
\item Ge detectors are under refurbishment, for deployment in early 2018.
\end{itemize}

  COHERENT deploys additional detectors for ancillary measurements (the ``in-COHERENT'' detectors).  Past detectors used to evaluate neutron backgrounda are SciBath~\cite{Cooper:2011kx} and a two-plane neutron scatter camera~\cite{Brennan:2009}; two detectors (the ``neutrino cubes'') are deployed to measure NINs in Fe and Pb.  The MARS detector~\cite{Roecker:2016juf} is currently measuring neutron backgrounds.  In addition to upgrades (additional mass) of NaI, LAr, and Ge, future possibilities include additional measurements of CC and inelastic NC events, a ``mini-HALO'' detector, and
a D$_2$O detector to reduce flux uncertainty using the well-known $\nu-d$ cross section. 
Ancillary measurements of quenching factors to reduce uncertainties will also be performed.

\section{Summary}
 
CEvNS, with an expected large,  $N^2$-dependent cross section, has been difficult to observe due to the tiny deposited energies of the nuclear recoils.  The interaction is accessible with low-energy threshold recoil detectors coupled with the ``sweet spot'' energies of stopped-pion neutrinos, accessible at the SNS.  
COHERENT has made the first measurement, and has harvested some of the lowest-hanging fruit of NSI interaction constraints.  More is to come, with upgrades of existing detectors, more
targets, and new ideas. 

\section*{Acknowledgments}

The author's COHERENT research is supported by the Department of Energy Office of High Energy Physics.  The COHERENT collaboration is supported by multiple sources; see ~\cite{Akimov:2017ade}.

\bibliographystyle{JHEP}
\bibliography{refs}

\end{document}